\documentclass[twocolumn,pra,showpacs]{revtex4}
\usepackage{graphics}

\begin{document}

\title{Optomechanical characterization of acoustic modes in a mirror}

\author{T. Briant}
\author{P.F. Cohadon}
\author{A. Heidmann}
\author{M. Pinard}
\affiliation{Laboratoire Kastler Brossel, Case 74, 4 place Jussieu, F75252
Paris Cedex 05, France}
\thanks{Unit\'{e} mixte de recherche du Centre National de la Recherche
Scientifique, de l'Ecole Normale Sup\'{e}rieure et de l'Universit\'{e} Pierre
et Marie Curie}
\homepage{www.spectro.jussieu.fr/Mesure}

\date{May 20, 2003}

\begin{abstract}
We present an experimental study of the internal mechanical vibration modes
of a mirror. We determine the frequency repartition of acoustic resonances
via a spectral analysis of the Brownian motion of the mirror, and the spatial
profile of the acoustic modes by monitoring their mechanical response to a
resonant radiation pressure force swept across the mirror surface. We have
applied this technique to mirrors with cylindrical and plano-convex
geometries, and compared the experimental results to theoretical predictions.
We have in particular observed the gaussian modes predicted for plano-convex
mirrors.
\end{abstract}

\pacs{42.50.Lc, 04.80.Nn, 05.40.Jc}

\maketitle

\section{Introduction}

Mirror thermal noise is a limitation for the sensitivity in many optical
precision measurements such as in gravitational-wave interferometers
\cite{saulson,brada,abra}, and is an important issue on the way to the
Standard Quantum Limit of displacement measurement \cite{caves,jeakel,bragi}.
Deformations of the mirror surface due to the thermal excitation of internal
acoustic modes strongly depend on the substrate geometry. The resulting noise
has been theoretically studied either by a decomposition of the motion over
the acoustic modes \cite{bondu,gillespie} or by a direct application of the
fluctuations-dissipation theorem \cite{hello,levin}.

Some experiments are currently under way in the gravitational-wave community,
both for the internal thermal noise \cite{LFF}, and for the pendulum thermal
noise \cite{Slagmolen,Leonhardt}. They are all of the ``direct approach''
kind: they aim at directly measuring the off-resonance thermal noise seen by
the light through the phase fluctuations of the field reflected by a cavity
with the mirror under study. Such tabletop experiments have to deal with two
difficulties: they have to work with a sufficient sensitivity in the
frequency band of interest, which explains why results are up to now only
available for the related photothermal noise
\cite{BraginskyPhoto,CerdonioPinard}, where the effect can be enhanced by
modulating the laser intensity \cite{DeRosa,Rao}. They must also use an
optical waist not too small as compared to the ones of kilometer-sized
interferometers \cite{brada,abra}. These two limitations are troublesome
since the transposition of the measured noise levels to more realistic
experimental conditions is far from trivial
\cite{saulson,Gonzalez,Kajima,Ohishi}.

We present here an experiment on the ``normal mode approach'' side. Our setup
\cite{hadjar,exp1,exp2} allows one to check for the accurate optomechanical
characteristics of every mechanical resonance of the mirror, including
individual quality factor and spatial profile. Such a detailed knowledge
rules out objections of the second kind. Applied to mirrors with a
cylindrical geometry, it should help in estimating the thermal noise level
sensed by the light in interferometers, where this particular geometry is
used. It also allows one to investigate the low-frequency optomechanical
behavior of mirror for quantum optics purposes, by summing the accurate
contributions of all acoustic modes. It is of great interest for mirrors with
the plano-convex geometry, where confined acoustic modes with a high quality
factor may enhance the coupling with light and could be a tool for the
experimental demonstration of quantum effects of radiation pressure
\cite{caves,jeakel,bragi,pinard}.

In Sec. \ref{internal} we remind some theoretical background on thermal noise
in the case of internal acoustic modes of a mirror. Sec. \ref{experiment} is
dedicated to the description of our experimental setup. Two different
geometries are then investigated: cylindrical mirrors (Sec. \ref{cyl}) and
plano-convex mirrors (Sec. \ref{pc}).

\section{Internal modes of a mirror}
\label{internal}

A light beam reflected on a moving mirror experiences a phase shift
proportional to the mirror displacement. One can take advantage of this
effect to detect displacements of the mirror with a very high sensitivity.
These displacements may be due to the propagation of acoustic waves in the
substrate which induces a deformation of its surface, or to a global
displacement of the mirror, due for example to the excitation of pendulum
modes for a suspended mirror. Both internal
\cite{bondu,hadjar,gillespie,exp1,exp2} and external \cite{tittonen,dorsel}
modes have an equivalent effect on the phase shift of the light reflected by
the mirror. Since external degrees of freedom strongly depend on the clamping
of the mirror, they are not considered in this paper. They usually have low
resonance frequencies and can easily be experimentally distinguished from
internal modes.

The propagation of the internal deformation ${\bf u}\left( {\bf r},t\right) $
in the mirror bulk is ruled by the elasticity equation \cite{elastic},
\begin{equation}
\rho \frac{\partial^2}{\partial t^2}{\bf u}\left( {\bf r},t\right) =\mu
\Delta {\bf u}\left( {\bf r},t\right) +\left( \lambda+\mu\right) \nabla
\left( \nabla. {\bf u}\left( {\bf r},t\right) \right) , \label{propag}
\end{equation}
where $\rho$ is the density of the material, $\lambda$ and $\mu$ its Lam\'e
constants. One can search for solutions of the form ${\bf u}\left( {\bf
r},t\right) = {\bf u}\left( {\bf r}\right) e^{-i\Omega t}$, where $\Omega$ is
the evolution frequency of the acoustic wave. Solutions of eq. (\ref{propag})
satisfying boundary conditions exist only for discrete frequencies $ \Omega_n
$, and provide an orthogonal basis ${\bf u}_n\left( {\bf r}\right) $ of
normal modes of vibration.

Any displacement ${\bf u}\left( {\bf r},t\right) $ can be decomposed as a sum
over all modes,
\begin{equation}
{\bf u}\left( {\bf r},t\right) =\sum_n a_n\left( t\right) {\bf u}_n\left(
{\bf r}\right) . \label{modal}
\end{equation}
The amplitude $a_n\left( t\right) $ depends on the force applied on the
mirror. If ${\bf F}\left( {\bf r},t\right) $ denotes the force by unit
surface, the Hamilton equations give an evolution of $a_n\left( t\right) $
equivalent to the one of a forced harmonic oscillator \cite{pinard},
\begin{equation}
\frac{d^2}{dt^2} a_n\left( t\right) +\Omega_n^2 a_n\left( t\right) =
\frac{1}{M_n} \langle{\bf F}\left( {\bf r},t\right) ,{\bf u}_n\left( {\bf
r}\right) \rangle , \label{a_n}
\end{equation}
where $M_n$ is the effective mass of the mode, proportional to the volume of
the mode inside the mirror, and where the brackets stand for the spatial
overlap over the mirror surface,
\begin{equation}
\langle{\bf F}\left( {\bf r},t\right) ,{\bf u}_n\left( {\bf r}\right) \rangle
= \int_S d^2{\bf r}\ {\bf F}\left( {\bf r},t\right) .{\bf u}_n\left( {\bf
r}\right) . \label{overlap}
\end{equation}
The Fourier component $a_n\left[ \Omega\right] $ is deduced from eq.
(\ref{a_n}),
\begin{equation}
a_n\left[ \Omega\right] = \chi_n\left[ \Omega\right] \langle{\bf F}\left[
{\bf r},\Omega\right] ,{\bf u}_n\left( {\bf r}\right) \rangle , \label{a_nF}
\end{equation}
where $\chi_n$ is the effective susceptibility of mode $n$ defined as,
\begin{equation}
\chi_n\left[ \Omega\right] =\frac{1}{M_n\left( \Omega_n^2 -\Omega^2\right) }.
\label{chi_n}
\end{equation}
$\chi_n$ is similar to the susceptibility of a harmonic oscillator without
dissipation, of resonance frequency $\Omega_n$ and mass $M_n$. As shown by
eqs. (\ref{modal}) and (\ref{a_nF}), the motion of the mirror is the
superposition of the responses of harmonic oscillators forced by the spatial
overlap between the force applied on the mirror and the modes.

The coupling of the mirror with a thermal bath is described by Langevin
forces and dampings which appear as additional imaginary parts in the
mechanical susceptibilities $\chi_n$. These are then similar to the ones of
damped harmonic oscillators,
\begin{equation}
\chi_n\left[ \Omega\right] =\frac{1}{M_n\left( \Omega_n^2
-\Omega^2-i\Phi_n\left[ \Omega\right] \Omega_n^2 \right) }, \label{chi_n_eff}
\end{equation}
where $\Phi_n\left[ \Omega\right] $ is the loss angle of the mode, related to
the damping rate $\Gamma_n$ by,
\begin{equation}
\Phi_n\left[ \Omega\simeq\Omega_n\right] =\Gamma_n/\Omega_n.
\end{equation}
The evolution of the amplitude $a_n$ [eq. (\ref{a_nF})] now becomes,
\begin{equation}
a_n\left[ \Omega\right] = \chi_n\left[ \Omega\right] \left( \langle{\bf
F}\left[ {\bf r},\Omega\right] ,{\bf u}_n\left( {\bf r}\right) \rangle +
F_{T,n}\left[ \Omega\right] \right). \label{a_n_lang}
\end{equation}
The Langevin forces $F_{T,n}$ are independent one from each other and their
spectra $S_{T,n}$ are related by the fluctuations-dissipation theorem to the
dissipative part of the mechanical susceptibility \cite{landau},
\begin{equation}
S_{T,n}\left[ \Omega\right] =-\frac{2k_B T}{\Omega}Im\left( 1/\chi_n\left[
\Omega\right] \right) , \label{STn}
\end{equation}
where $k_B$ is the Boltzmann constant and $T$ the temperature of the thermal bath.
Since this noise spectrum is flat in frequency close to the resonance,
\begin{equation}
S_{T,n}\left[ \Omega\simeq\Omega_n\right] =2M_n \Gamma_n k_B T,
\end{equation}
one gets from eqs. (\ref{a_n_lang}) and (\ref{STn}) a displacement spectrum
due to mode $n$ which has a lorentzian shape of width $\Gamma_n$. The
resulting Brownian motion of the mirror is the superposition of the responses
of each acoustic mode to these Langevin forces and presents resonant peaks at
each resonance frequency.

\section{Experimental setup}
\label{experiment}

We first study the internal acoustic modes of a flat {\it Newport
SuperMirror} coated on a cylindrical substrate made of fused silica. The
mirror is $6.35\ mm$-thick with a diameter of $25.4\ mm$ and is clamped on
its edge at three $120^\circ$-shifted points. With a second concave mirror
(curvature radius of 1 $m$) coated on a similar substrate, it provides a
high-finesse symmetric Fabry-Perot cavity. The cavity is $0.23\ mm$ long and
has an optical finesse of 26000.

\begin{figure}
\includegraphics{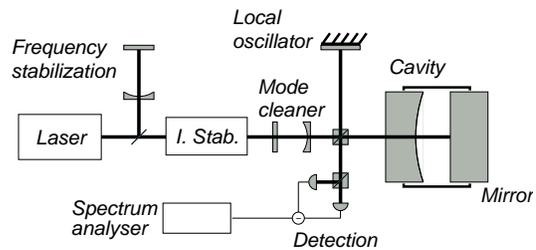}
\caption{Experimental setup. A cylindrical mirror is used as the end mirror
of a high-finesse cavity. A frequency and intensity-stabilized laser beam is
sent into the cavity, and the phase of the reflected beam is measured by a
homodyne detection. A spectrum analyzer monitors the power of phase
fluctuations of the reflected beam. } \label{setup1}
\end{figure}

The optical setup \cite{hadjar} is shown in Fig. \ref{setup1}. A $1\ mW$
frequency and intensity-stabilized light beam provided by a Titane-Sapphire
laser working at $810\ nm$ is sent into the cavity. A mode cleaner also
ensures a spatial filtering of the laser beam. A homodyne detection monitors
the phase of the field reflected by the cavity.

For a resonant cavity, the detected signal reflects the global phase-shift
experienced by the intracavity field, which can be written as,
\begin{equation}
\psi\left( t\right) =4\pi\frac{\widehat u\left( t\right) }{\lambda},
\label{phase}
\end{equation}
where $\lambda$ is the optical wavelength, and $\widehat u\left( t\right) $
is the longitudinal mirror displacement averaged over the beam profile
\cite{bondu,gillespie,pinard}. The field of the fundamental mode in the
cavity is characterized by a transverse gaussian structure $v_0\left( {\bf
r}\right) $ at the mirror surface given by,
\begin{equation}
v_0\left( {\bf r}\right) =\frac{\sqrt{2/\pi}}{w_0}e^{-{\bf r}^2/{w_0^2}},
\label{structure}
\end{equation}
where $w_0$ is the optical waist. The averaged displacement
$\widehat{u}\left( t\right) $ corresponds to the spatial overlap between the
longitudinal mirror displacement $u\left( {\bf r},t\right) $ and the
intensity profile $v_0^2\left( {\bf r}\right) $ of the light beam,
\begin{equation}
\widehat{u}\left( t\right) =\langle u\left( {\bf r},t\right) ,v_0^2\left(
{\bf r}\right) \rangle. \label{u_chapeau}
\end{equation}

When no external force is applied on the mirror, the measurement gives the
noise spectrum $S_{\hat{u}}\left[ \Omega\right] $ of the displacement due to
the thermal noise. From eqs. (\ref{modal}) and (\ref{chi_n_eff}) to
(\ref{STn}) one gets,
\begin{equation}
S_{\hat{u}}\left[ \Omega\right] =\sum_n \frac{2\Phi_n\Omega_n^2 k_B T}{\Omega
M_n^{\rm eff}} \frac{1}{\left( \Omega_n^2 -\Omega^2\right)
^2+\Phi_n^2\Omega_n^4}, \label{S_u}
\end{equation}
where the effective mass $M_n^{\rm eff}$ is defined by,
\begin{equation}
M_n^{\rm eff}=\frac{M_n}{\left| \langle u_n\left( {\bf r}\right) ,v_0^2\left(
{\bf r}\right) \rangle \right| ^2}. \label{M_n^eff}
\end{equation}
For acoustic modes with a characteristic size much larger than the optical
waist $w_0$, the spatial overlap in eq. (\ref{M_n^eff}) can be approximated
to the longitudinal displacement $u_n\left( {\bf r}=0\right) $ at the center
of the mirror.

Equation (\ref{S_u}) shows that the thermal noise spectrum probed by the
optomechanical sensor is equivalent to the one of a set of independent
harmonic oscillators in thermal equilibrium at temperature $T$, with
effective masses $M_n^{\rm eff}$ related to the coupling between the acoustic
modes and the light beam. In particular, for a light beam perfectly centered
on the mirror, a mode with a null displacement at the center has an infinite
effective mass and does not contribute to the signal.

\begin{figure}
\includegraphics{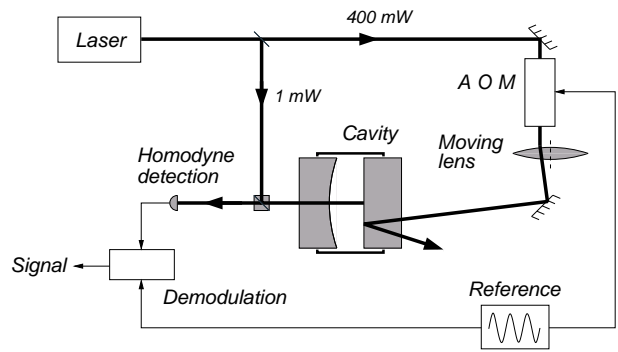}
\caption{A 400mW laser beam, intensity-modulated by an acousto-optic
modulator (AOM), is sent from the back on the mirror. A moving lens deflects
the beam in the horizontal and vertical planes to scan the whole surface of
the mirror. A reference signal synchronizes the AOM and the detection in
order to extract the amplitude of the mirror response to the radiation
pressure force.} \label{manip-spatiale}
\end{figure}

Optomechanical parameters such as the resonance frequencies $\Omega_n$, the
widthes $\Gamma_n$, and the effective masses $M_n^{\rm eff}$ can be deduced
from the observation of the thermal noise spectrum. It gives no information
on the spatial structure of the acoustic modes since the noise spectrum is
only sensitive to the displacements at the center of the mirror. The spatial
profile can be probed by the response of the mirror to an external force. The
experimental setup is then modified as shown in Fig. \ref{manip-spatiale}. An
intense auxiliary laser beam ($400\ mW$) is reflected from the back on the
mirror. Its intensity is modulated at the resonance frequency $\Omega_n$ of a
given mode by an acousto-optic modulator. The beam spot can be deflected in
the horizontal and vertical planes by a lens mounted on two motorized
micrometric translations. This provides a confined radiation pressure force
sweepable over the mirror surface. A demodulation system synchronized with
the intensity modulation extracts the amplitude and phase of the mechanical
response of the mirror \cite{exp2}.

According to eq. (\ref{a_n_lang}) the mechanical response is proportional to
the overlap of the force with the spatial distribution ${\bf u}_n\left( {\bf
r}\right) $ of the mode. The auxiliary laser beam is focused on the mirror
surface with a waist of 100 $\mu m$. The radiation pressure force ${\bf
F}\left( {\bf r}\right) $ is thus localized over dimensions small compared to
the characteristic size of the modes, and the overlap can be approximated to,
\begin{equation}
\langle{\bf F}\left[ {\bf r},\Omega_n\right] ,{\bf u}_n\left( {\bf r}\right)
\rangle =F \left[ \Omega_n\right] u_n\left( {\bf r}_0\right) , \label{force}
\end{equation}
where ${u}_n\left( {\bf r}_0\right) $ is the longitudinal displacement at the
position ${\bf r}_0$ of the beam spot on the mirror surface, and $F \left[
\Omega_n\right] $ is the amplitude of the radiation pressure force. The
displacement measured with the homodyne detection is thus proportional to the
spatial structure ${u}_n\left( {\bf r}_0\right) $ at point ${\bf r}_0$.

The motion of the lens is computed in order to move the beam spot along 50
horizontal lines, uniformly distributed along the vertical direction over the
mirror surface. For each line, the beam spot is continuously swept from one
edge of the mirror to the other one and the amplitude of the displacement
modulation is acquired with a digital oscilloscope. The speed of the spot
over the surface is set to $5\ mm/s$. Given the time constant of the
mechanical response, this speed is slow enough to ensure a spatial resolution
better than half a millimeter, way below the expected spatial wavelength of
the acoustic modes.

\section{cylindrical mirror}
\label{cyl}

We present in this section the characterization of acoustic modes observed
for cylindrical mirrors, and we compare the experimental results to a
theoretical model developed for this particular geometry \cite{hutchi} and
used in the framework of gravitational-wave detection \cite{bondu,gillespie}.

\begin{figure}
\scalebox{0.6}{\includegraphics{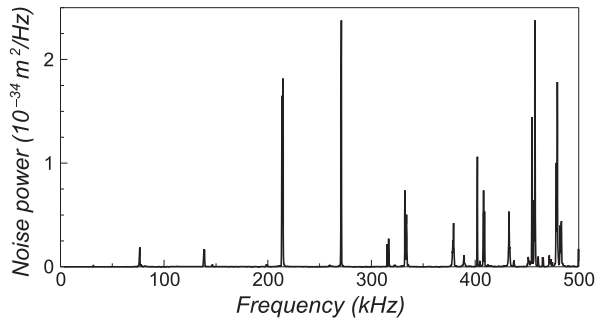}}
\caption{Displacement noise $S_{\hat{u}}\left[ \Omega\right] $ from 10 to
$500\ kHz$. Each peak corresponds to the thermal noise resonance associated
with an acoustic mode of one mirror of the cavity.} \label{spectre}
\end{figure}

Figure \ref{spectre} shows the phase-noise spectrum of the reflected beam
acquired between $10$ and $500\ kHz$ with a resolution bandwidth of $300\
Hz$. This spectrum reflects the cavity length fluctuations due to the
Brownian motion of both mirrors at room temperature. We have performed a
calibration of the measured displacements by using a frequency modulation of
the laser beam. According to eq. (\ref{phase}), such a modulation changes the
length reference given by the wavelength $\lambda$ of the laser. A modulation
$\Delta \nu$ of the optical frequency $\nu$ is thus equivalent to a
displacement $\Delta u$ related to $\Delta \nu$ by,
\begin{equation}
\frac{\Delta \nu}{\nu}=\frac{\Delta u}{L}, \label{cal}
\end{equation}
where $L$ is the cavity length. We have applied a frequency modulation of $7\
kHz$ with a carrier frequency swept over the whole frequency range of
interest (from $10\ kHz$ up to a few megahertz). Monitoring the phase
modulation of the reflected beam, equation (\ref{cal}) allows one to
calibrate in $m/\sqrt{Hz}$ the measured displacement of the mirror. As shown
in Fig. \ref{spectre}, the thermal displacements are of the order of
$10^{-17}\ m/\sqrt{Hz}$, way above the quantum-limited sensitivity, of the
order of $10^{-19}\ m/\sqrt{Hz}$ \cite{hadjar}.

The phase-noise spectrum exhibits sharp peaks at each resonance frequency of
internal acoustic modes. Their spectral repartition depends on the geometry
of the mirror and cannot be analytically computed in the general case. For a
cylinder, however, a specific method has been proposed by Hutchinson
\cite{hutchi}, and we use the software {\sc Cypres} developed by Bondu and
Vinet to determine the optomechanical characteristics of the {\sc Virgo}
mirrors \cite{bondu}. Acoustic modes are classified according to a
circumferential order $n$ (corresponding to the angular geometry), a parity
$\xi$ (equal to 0 for modes having both faces vibrating in phase, and to 1
for those where faces vibrate with opposite phases), and an order number $m$
(related to the radial structure).

\begin{figure}
\scalebox{0.6}{\includegraphics{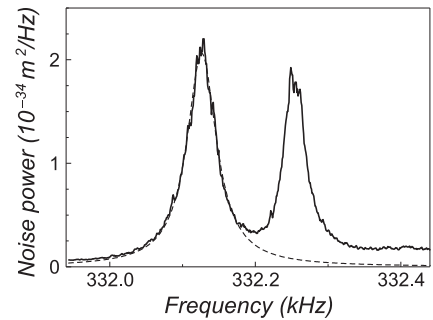}}
\caption{Thermal noise spectrum around a resonance frequency (mode $n\xi m=0\
1\ 3$). The twin-peaks structure corresponds to the resonances of the two
mirrors of the cavity. The dashed line is a Lorentzian fit of the first
peak.} \label{spectre2}
\end{figure}

The computation with {\sc Cypres} for a $6.35\ mm$-thick cylinder with a
diameter of $25.4\ mm$ provides a set of frequencies, allowing us to label
every peak observed in the experimental thermal spectrum of Fig.
\ref{spectre}. Table \ref{table} presents the comparison between the
theoretical resonance frequencies (second column) and the experimental ones
(third column). Discrepancies between experimental and theoretical values are
less than 2\%, of the same order of magnitude as the uncertainty on the
mirror dimensions. The observation of modes with a non-zero circumferential
order $n$, which have no displacement at the center, shows that the laser
beam is not perfectly centered on the mirror. This is due to a slight
misparallelism between the mirrors of the cavity.

\begin{table}
\scriptsize
\begin{tabular}[t]{|c|c|c|}
\hline
Mode & {\sc Cypres} & Exp \\
$n\xi m$ & $(kHz)$ & $(kHz)$ \\
\hline \hline
0 0 1 & 143 & 146 \\
0 0 2 & 377 & 379 \\
0 0 3 & 405 &     \\
0 0 4 & 460 & 457 \\
0 0 5 & 468 & 465 \\
0 0 6 & 483 & 483 \\
\cline{1-3}
0 1 1 & 73  & 75  \\
0 1 2 & 210 & 213 \\
0 1 3 & 330 & 332 \\
0 1 4 & 406 & 408 \\
0 1 5 & 491 & 494 \\
\hline
\end{tabular}
\ \
\begin{tabular}[t]{|c|c|c|}
\hline
Mode & {\sc Cypres} & Exp \\
$n\xi m$ & $(kHz)$ & $(kHz)$ \\
\hline \hline
1 0 5 & 435 & 432 \\
\hline
1 1 1 & 135 & 138 \\
1 1 2 & 268 & 270 \\
1 1 3 & 317 & 316 \\
1 1 4 & 334 &     \\
1 1 5 & 400 & 402 \\
1 1 6 & 436 & 437 \\
1 1 7 & 475 & 478 \\
\hline
2 1 3 & 314 & 315 \\
2 1 6 & 459 & 460 \\
\hline
\end{tabular}
\caption{Resonance frequencies of the acoustic modes for a cylindrical
mirror. Second column: theoretical results obtained with {\sc Cypres}. Third
column: experimental results.} \label{table}
\end{table}

Close to a particular resonance frequency, the mirror motion is mainly ruled
by the resonant mode. Figure \ref{spectre2} presents the phase-noise spectrum
around $332\ kHz$ with a frequency span reduced to $500\ Hz$ and a resolution
bandwidth of $3\ Hz$. This small resolution bandwidth allows one to resolve
the resonance structure, unlike in Fig. \ref{spectre} where the wider
bandwidth broadens and lowers the resonances. The coupling mirror has a
concave side, but its curvature radius is much larger than every other
dimension so that it can be considered as a cylinder. Both mirrors have thus
similar acoustic modes and the spectrum has a twin-peaks structure. The
discrepancy between the resonance frequencies of the mirrors is less than 1\%
for every resonance observed in Fig. \ref{spectre}.

According to eq. (\ref{S_u}), the thermal noise spectrum has a lorentzian
shape centered at the resonance frequency with a width equal to the
relaxation rate $\Gamma_n$ and an area proportional to $k_BT/M_n^{\rm eff}
\Omega_n^2$. The dashed line in Fig. \ref{spectre2} is a lorentzian fit of
the first peak. It gives a mechanical quality factor $Q_n=\Omega_n/\Gamma_n$
equal to $6\,600$ for this mode. The quality factor strongly depends on the
spatial expansion of the acoustic mode and on the clamping of the mirror.
With our three-points clamping, we have obtained quality factors from
$1\,400$ up to $28\,000$ for the various modes presented in Table
\ref{table}.

The effective mass $M_n^{\rm eff}$ can be deduced from the area of the
lorentzian fit. For modes of Table \ref{table}, one gets masses at least 10
times larger than the theoretical ones. This seems to be due to the stress
induced by the three-points clamping, while the mirror is assumed to be free
in the theoretical derivation. Note also the absence of some modes in Table
\ref{table}. These modes have a very high theoretical mass (550 $g$ for mode
$0\ 0\ 3$), whereas the typical mass for other modes is a fraction of gram.
Their thermal resonances have thus not been seen in the noise spectrum.

\begin{figure}
\scalebox{0.48}{\includegraphics{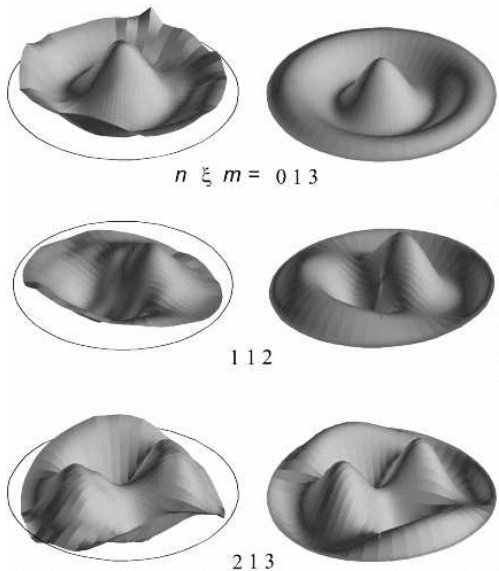}}
\caption{Spatial profiles of modes $n\xi m=0\ 1\ 3$ (top), 1 1 2 (middle),
and 2 1 3 (bottom). Left: polar plots of experimental response to the
external force. Solid circles represent the mirror edge. Right: theoretical
profiles obtained with {\sc Cypres}.} \label{modes}
\end{figure}

We have studied the spatial distribution of each mode observed in the thermal
noise spectrum of Fig. \ref{spectre}, by using the sweepable auxiliary laser
beam described in the previous section (Fig. \ref{manip-spatiale}). For each
mode, the modulation frequency is set to the acoustic resonance frequency,
and the beam spot is scanned over the mirror surface. The amplitude of phase
modulation of the reflected beam is stored in a computer and plotted in polar
coordinates as a function of the beam-spot position. The left part of Fig.
\ref{modes} shows the result for three modes ($n\xi m=0\ 1\ 3$, $1\ 1\ 2$,
and $2\ 1\ 3$). The solid circle represents the mirror edge. The vertical
scale on these plots corresponds to the observed displacements which are of
the order of $10^{-15}\ m$. The right part of Fig. \ref{modes} shows the
theoretical profiles computed with {\sc Cypres}. We have obtained a very good
agreement between the experimental and theoretical spatial distributions, for
every mode presented in Table \ref{table}. This is, to our knowledge, the
first full experimental demonstration of Hutchinson's theory.

\section{plano-convex mirror}
\label{pc}

In this section we study the acoustic modes of a mirror with a plano-convex
geometry. For this particular shape, in the framework of a paraxial
approximation, the wave propagation equation (\ref{propag}) has solutions
with a gaussian spatial structure \cite{wilson,zarka,pinard}. These modes are
confined at the center of the mirror and may have quality factors independent
of the mirror clamping.

The end mirror of the cavity is replaced by a mirror coated on the flat side
of a plano-convex resonator with a diameter of $34\ mm$, a curvature radius
$R$ of $150\ mm$ for the convex side, and a thickness $h_0$ at the center of
$2.65\ mm$. Hutchinson's theory is no longer valid for this geometry. We have
thus used a finite elements model, based on the ProMechanica software, to
determine the resonance frequencies and spatial structures of the acoustic
modes. The spatial profile and resonance frequency of the acoustic modes vary
continuously when the curvature radius $R$ of the convex side varies from
infinity (cylindrical mirror) to a finite value (plano-convex mirror). It is
then possible to label the acoustic modes of a plano-convex mirror with the
same parameters $n$, $\xi$ and $m$. As long as the curvature radius is much
larger than the diameter, the discrepancy of frequencies between a
cylindrical and a plano-convex mirror is less than a few percents.

\begin{figure}
\scalebox{0.48}{\includegraphics{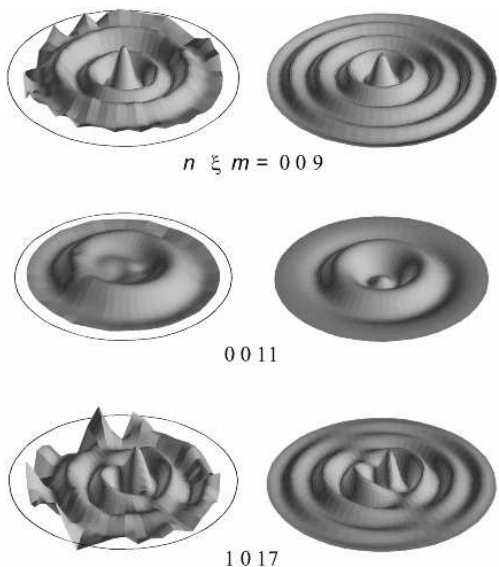}}
\caption{Spatial profiles of modes representative of a nearly cylindrical
geometry, obtained with a plano-convex mirror. The three modes $n\xi m=$ 0 0
9 (top), 0 0 11 (middle), and 1 0 17 (bottom) have a resonance frequency
around 1.1 $MHz$. Left: experimental results. Right: theoretical counterparts
computed with {\sc Cypres} for a cylindrical mirror of same dimensions.}
\label{modes_plein}
\end{figure}

First column in Fig. \ref{modes_plein} shows the experimental profiles of
three acoustic modes observed at $1158$, $1194$ and $1118\ kHz$. The second
column presents the theoretical profiles computed with {\sc Cypres} for a
cylindrical mirror of same dimensions. Theoretical resonance frequencies are
respectively 1159, 1195, and 1100 $kHz$. Experimental frequencies and spatial
structures are both in very good agreement with theoretical results. These
modes present a spatial expansion over the whole surface of the mirror,
either via a slowly varying radial profile (mode $0\ 0\ 11$), or via multiple
oscillations (modes $0\ 0\ 9$ and $1\ 0\ 17$). Quality factors for these
modes are of the order of only a few tens of thousands.

\begin{figure}
\scalebox{0.48}{\includegraphics{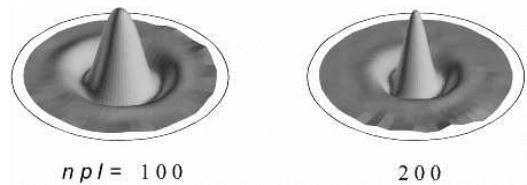}}
\caption{Experimental spatial profiles of gaussian modes $npl=$ 1 0 0 at 1.1
$MHz$ (left), and 2 0 0 at 2.1 $MHz$ (right), obtained with a plano-convex
mirror.} \label{modes_gauss_plein}
\end{figure}

In parallel with these modes representative of a nearly cylindrical geometry,
we have observed confined acoustic modes which are not predicted by
Hutchinson's theory (see Fig. \ref{modes_gauss_plein}). These modes are
specific to the plano-convex geometry and can be deduced from a paraxial
approximation valid when the curvature radius of the convex side is large
compared to the mirror thickness. The propagation equation (\ref{propag}) has
solutions similar to the gaussian optical modes of a Fabry-Perot cavity
\cite{kogel}. Within this approximation, shear modes have no longitudinal
displacement, and the light beam is only sensitive to compression modes.
These modes are labelled by three indexes $n$, $p$, and $l$, and their
longitudinal displacements at a point $(r,\theta,z)$ inside the substrate are
given by \cite{pinard,wilson,zarka},
\begin{eqnarray}
u_{n p l}\left( r,\theta,z\right) &=& e^{-r^2/w_n^2}\left(
\frac{r}{w_n}\right) ^lL_p^l\left( \frac{2r^2}{w_n^2}\right)
\times \nonumber\\
& &\cos\left( l\theta\right) \cos \left( \frac{n\pi}{h\left( r\right)
}z\right) , \label{gauss}
\end{eqnarray}
where $L^l_p$ is the Laguerre polynomial, $h\left( r\right) $ the thickness
at radial position $r$ (equal to $h_0$ for $r=0$), and $w_n$ the acoustic
waist defined by,
\begin{equation}
w_n^2=\frac{2h_0}{n\pi}\sqrt{Rh_0}.
\label{waist}
\end{equation}
The resonance frequencies $\Omega_{npl}$ are given by,
\begin{equation}
\Omega_{npl}^2= \left( \frac{\pi c_l}{h_0}\right) ^2\left[
n^2+\frac{2}{\pi}\sqrt{\frac{h_0}{R}}n\left( 2p+l+1\right) \right] ,
\label{freq_gauss}
\end{equation}
where $c_l$ is the longitudinal sound velocity. Gaussian modes correspond to
series of overtones associated with the longitudinal index $n$. For each
overtone, numbers $p$ and $l$ define the transverse order.

The parameters of our plano-convex mirror lead to a fundamental frequency of
$1171\ kHz$. The experimental thermal noise spectrum shows two huge
resonances at $1116$ and $2121\ kHz$. Figure \ref{modes_gauss_plein} presents
the experimental spatial profile of these two modes, which fit well with the
fundamental modes of the first two overtones (modes $npl = 1\ 0\ 0$ and $2\
0\ 0$). Half-amplitude widthes of the radial profiles give an acoustic waist
$w_1 = 3.5\ mm$ for the first mode, and $w_2= 2.4 \ mm$ for the second,
somewhat smaller than the expected values (5.8 and 4.1 $mm$, respectively).
The ratio $w_1/w_2$ is however very close to $\sqrt2$, in agreement with eq.
(\ref{waist}).

\begin{figure*}
\scalebox{0.37}{\includegraphics{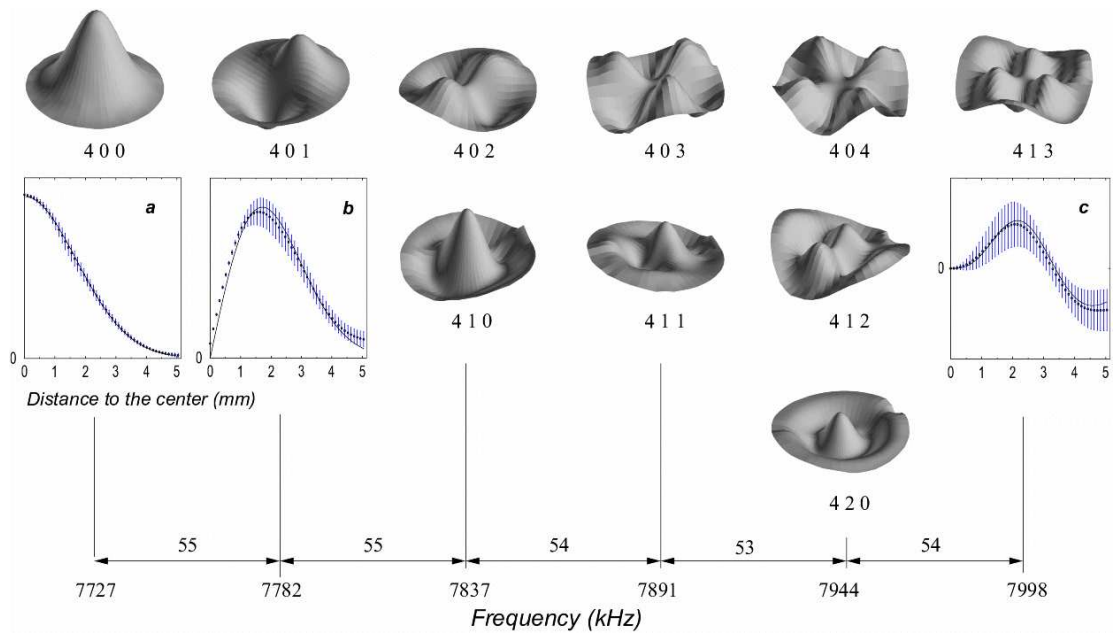}}
\caption{Experimental spatial profiles of the fourth overtone gaussian modes
($n=4$) of a plano-convex mirror. Modes with the same value of $2p+l$ present
the expected frequency degeneracy. Inserts $a$, $b$ and $c$ show the radial
profiles of modes $npl=$ 4 0 0, 4 0 1, and 4 1 3, respectively, as a function
of the radial distance $r$ to the center.} \label{mode_gauss}
\end{figure*}

The corresponding peaks in the thermal noise spectrum have a width of a few
hertz, at the resolution limit of our spectrum analyzer. We have thus
measured the mechanical quality factors $Q_n$ of both modes
 by monitoring the response time $1/\Gamma_n$ of the mirror to
an external impulsive force resonant with the mode \cite{exp1}. We have
obtained quality factors of respectively $350\,000$ and $650\,000$ for these
modes. They appear to be much larger than the ones obtained for non-confined
acoustic modes. The effective masses of both modes are respectively $40$ and
$50\ mg$, below the expected values ($153$ and $77\ mg$) \cite{pinard}, and
more than 10 times below the experimental effective masses obtained for
non-gaussian modes (360 $mg$ for mode $n\xi m=0\ 0\ 9$, and 1550 $mg$ for
mode $0\ 0\ 11$).

These results can be compared to previous results obtained by x-ray
topography with plano-convex resonators \cite{wilson,zarka}. In these
experiments, the resonator is excited by piezoelectric effect with electrodes
deposited on both sides. In our experiment we excite the acoustic modes by a
radiation pressure applied on only one side. We thus detect modes of odd
overtone as well as those of even overtone, whereas only odd overtones are
observed by x-ray topography.

As already observed in Ref. \cite{zarka}, the experimental profiles fit
better and better with the theoretical ones as the overtone number $n$
increases and the modes are more confined [see eq. (\ref{waist})]. The
fundamental mode observed in Ref. \cite{zarka} presents a very noisy profile,
whereas modes of third and fifth overtones fit well with a gaussian profile.
Figure \ref{mode_gauss} presents the spatial distributions we obtained for
the gaussian modes of the fourth overtone. The plano-convex mirror now has a
diameter of $12\ mm$, a curvature radius of $180\ mm$, and a thickness of
1.55 $mm$. The transverse fundamental mode ($npl=4\ 0\ 0$) has a resonance
frequency of 7727 $kHz$, in excellent agreement with the theoretical value
(7747 $kHz$). The transverse indexes $p$ and $l$ of a mode can be deduced
from the number of zeroes of the radial function and from the cylindrical
symmetry of the spatial distribution. This classification shows that the
transverse modes have the expected degeneracy. For example, modes $4\ 2\ 0$,
$4\ 1\ 2$, and $4\ 0\ 4$ have equal frequencies, with an agreement better
than 0.5\%. The frequency gap between non-degenerate modes is equal to $55\
kHz$, in good agreement with the theoretical value ($57\ kHz$).

Knowing the circumferential order $l$ of a mode, the radial function can be
extracted from the experimental distribution and compared to the theoretical
radial dependency given by eq. (\ref{gauss}). Inserts $a$, $b$, and $c$ in
Fig. \ref{mode_gauss} show the radial profile of different modes. Each
experimental point is the angular average of the experimental distribution at
a distance $r$ from the center, divided by the angular dependence $\cos\left(
l\theta\right) $. A gaussian fit of the 4 0 0 mode profile, shown by the
solid curve in insert $a$, gives an acoustic waist of $2.4\ mm$, slightly
larger than the $2.0\ mm$ expected. The other profiles in Fig.
\ref{mode_gauss} are compared with the theoretical radial function (solid
lines) using the same experimental value of the waist, without any adjustable
parameter.

Finally note that our resonator is made of an isotropic material and has the
cylindrical symmetry. We thus find that gaussian modes are well described by
the Laguerre polynomials basis, in contrast with the acoustic modes of a
vibrating anisotropic plano-convex quartz crystal which present the Hermite
polynomials structure \cite{wilson}.

\section{Conclusion}
We have performed a high-sensitivity measurement of the optomechanical
properties of a mirror. Resonance peaks in the thermal noise spectrum of the
mirror displacement allow us to determine the frequencies and quality factors
of the acoustic modes of the substrate. We have presented a new technique of
spatial analysis of the acoustic modes, using the mirror response to a
radiation pressure force. Applied to a cylindrical mirror, this method gives
the resonance frequencies and spatial profiles of the acoustic modes, in very
good agreement with the theoretical model.

The study of a plano-convex mirror with a large curvature radius reveals a
similar behavior, with the same acoustic mode structure as a cylindrical
mirror. We have also detected confined modes with a gaussian structure, high
mechanical quality factors, and small effective masses. The experimental
observation of both kinds of acoustic modes gives a more detailed
characterization of the behavior of mirrors with a plano-convex geometry.

These results are of great interest for a better understanding of the
mechanical properties of a mirror substrate, either in the framework of
gravitational-wave detection where the internal thermal noise of the mirror
is a critical issue, or for the demonstration of quantum effects in the
optomechanical coupling between light and mirrors.

\acknowledgments We gratefully acknowledge Jean-Marie Mackowski for the
optical coating of the mirrors, Fran\c{c}ois Bondu and Jean-Yves Vinet for
stimulating discussions and for the {\sc Cypres} program, Luca Taffarello and
Olivier Arcizet for helping us with the finite elements computations.

\end{document}